\documentclass[aps,prl,reprint,amsmath,amssymb,amsfonts,floatfix, showpacs,intlimits, euspript,superscriptaddress]{revtex4-1}

\usepackage{graphicx}
\usepackage{dcolumn}
\usepackage{bm}
\usepackage{hyperref}
\usepackage{xcolor}
\hypersetup{
	colorlinks=true,
	linkcolor=blue,
	citecolor=blue,
	urlcolor=cyan,
}

\usepackage{xcolor}

\newcommand\numberthis{\addtocounter{equation}{1}\tag{\theequation}}

\begin{document}
	
	\preprint{APS/123-QED}
	
	\title{Femtosecond laser pulse driven caustic spin wave beams}
	
	\author{S. Muralidhar}
	\email{shreyas.muralidhar@physics.gu.se}
	\affiliation{Physics Department, University of Gothenburg, 412 96 Gothenburg, Sweden.}
	\author{R. Khymyn}
	\affiliation{Physics Department, University of Gothenburg, 412 96 Gothenburg, Sweden.}
	\author{A. A. Awad}
	\affiliation{Physics Department, University of Gothenburg, 412 96 Gothenburg, Sweden.}
	\author{A. Alem\'an}
	\affiliation{Physics Department, University of Gothenburg, 412 96 Gothenburg, Sweden.}
	\author{D. Hanstorp}
	\affiliation{Physics Department, University of Gothenburg, 412 96 Gothenburg, Sweden.}
	\author{J. \AA kerman}
	\email{johan.akerman@physics.gu.se}
	\affiliation{Physics Department, University of Gothenburg, 412 96 Gothenburg, Sweden.}
	\affiliation{Materials and Nano Physics, School of Engineering Sciences, KTH Royal Institute of Technology, Electrum 229, 164 40 Kista, Sweden.}
	
	
	\date{\today}
	
	\begin{abstract}
		Controlling the directionality of  spin waves is a key ingredient in wave-based computing methods such as magnonics. In this paper we demonstrate this particular aspect by using an all-optical point-like source of continuous spin waves based on frequency comb rapid demagnetization. The emitted spin waves contain a range of k-vectors and by detuning the applied magnetic field slightly off the ferromagnetic resonance (FMR), we observe X-shaped caustic spin wave patterns at 70$^{\circ}$ propagation angles as predicted by theory. When the harmonic of the light source approaches the FMR, the caustic pattern gives way to uniaxial spin wave propagation perpendicular to the in-plane component of the applied field. This field-controlled propagation pattern and directionality of optically emitted short-wavelength spin waves provide additional degrees of freedom when designing magnonic devices. 
	\end{abstract}
	
	\maketitle
	
	Caustics are widely known from optics as the envelopes of light rays refracted by curved surfaces or objects \cite{stavroudis2012optics, nye1999natural}. The notion, however, can be easily extended to any propagation medium whose anisotropy creates regions with enhanced amplitude of the wave field \cite{kravtsov2012caustics}. The phenomenon of caustic patterns encompasses an extraordinarily wide range of fields in physics, from understanding the physics of the universe, its creation, and dark matter dynamics \cite{Arnold1982}, to studying light patterns at the bottom of a pool due to surface curvature variations. Caustics also find applied uses such as the shadow optical method to study stress intensity factors in transparent materials \cite{Manogg1964,Theocaris1970,Theocaris1981,Kalthoff1987}.
	
	The important feature, known from the simple paraxial case in geometrical optics, is the structural stability of the caustics, which imply small fuzziness of the caustic beam under variation of the scattering surface curvature \cite{nye1999natural}. This opens a way to create non-diffractive beams in anisotropic continuous media, \emph{e.g.}~self-focusing beams of propagating phonons \cite{Taylor1969,Every1986,Maznev1996}. Similar effects have also been observed for spin waves (SWs) in magnetic materials, where the anisotropic properties can furthermore be readily modified by the strength and direction of the applied magnetic field. Early experiments to study self-focusing and caustic patterns were conducted on YIG films where the SWs were excited by microwave antennas  \cite{Bauer1998,Buttner2000,Schneider2010}. Later, SWs excited by RF antennas on structured NiFe waveguides showed caustic patterns where the waveguides were connected to an extended NiFe film  \cite{Demidov2009}. Detailed theoretical analyses of SW caustics is now available for ferromagnetic metals \cite{Veerakumar_PRB_2006}, insulators \cite{Veerakumar_IEEE_2006}, and antiferromagnetic thin films  \cite{Veerakumar2010}.  
	
	\begin{figure}
		\begin{center}
			\includegraphics[width=0.95\linewidth]{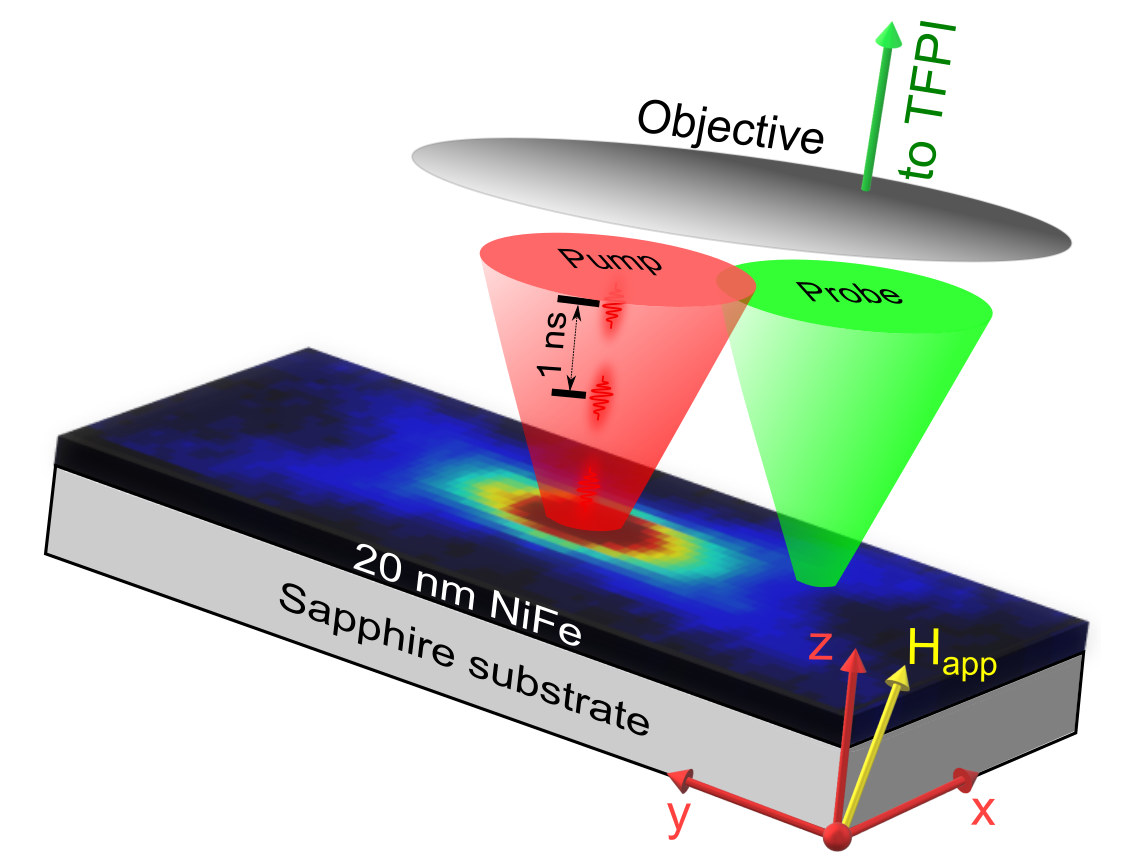}
			\caption{Schematic of the experiment.
				A 20 nm NiFe film on a transparent Sapphire substrate is pumped with a 1 GHz-pulsed femtosecond laser (120 fs) at a wavelength of  816 nm (red) and probed by a continuous single-mode 532 nm Brillouin Light Scattering laser (green). The magnetic field $H_{app}$ is applied at an oblique angle of 82$^\circ$ out-of-plane with its in-plane component along the y axis. Back-scattered photons (green arrow) are analysed using a 6-pass Tandem Fabry-Pérot Interferometer (TFPI).}
			\label{fig:1}
		\end{center}
	\end{figure}
	
	While all experimental observations of SW caustics have been based on SW emission using RF antennas, recent interest in exciting SWs using femtosecond lasers has rapidly increased \cite{VanKampen2002,Au2013,Yun2015,Iihama2016b,Kamimaki2017,Jackl2017,Savochkin2017,awad2019stimulated}.  However, no such optical SW excitation studies have shown any sign of caustics, which is likely due  to the inability of exciting sustained SWs in a focused region close to the diffraction limit of the light, which is required to excite SWs with high wavevectors, where caustics can be formed. Here we overcome this limitation by utilizing a 1 GHz repetition-rate femtosecond laser comb capable of sustaining continuous SWs \cite{Jackl2017,awad2019stimulated,Aleman_inst}. The emitted SWs contain a range of $k$-vectors and by slight detuning of the applied magnetic field off the ferromagnetic resonance (FMR), we observe X-shaped caustic SW patterns at around 70$^{\circ}$ angles as predicted by theory. When the harmonic of the light source approaches the FMR, the caustic pattern gives way to uniaxial SW propagation perpendicular to the in-plane component of the applied field.
	This field-controlled propagation pattern and directionality of optically excited short-wavelength SWs provide additional degrees of freedom when designing magnonic devices. The demonstrated SW caustics can \emph{e.g.}~be utilized as a potential signal splitter in magnonic networks   \cite{Heussner2017} without any requirements for additional patterning for such functions. 
	
	\emph{Experimental method:} Fig.~\ref{fig:1} shows the schematic of the experiment. A more in-depth description can be found in Ref. \onlinecite{Aleman_inst}. Briefly, the pump-probe experiment uses a mode-locked femtosecond laser operating at 816~nm and a repetition rate of 1 GHz with 120~fs long pulses to induce rapid demagnetization. Each laser pulse instantaneously increases the temperature of the electron gas in the thin metallic film. The thermal energy of the electrons is then rapidly transferred to the magnetic subsystem, creating a rapid demagnetization of the irradiated region and hence a rapid pulse of the local effective magnetic field. The periodic modulation of the demagnetizing field excites magnons with frequencies that are a multiple of the 1 GHz repetition rate of the pump-laser pulses \cite{awad2019stimulated}.  
	
	The excited SWs are studied using Brillouin Light Scattering (BLS) microscopy. A 532~nm continuous wave laser is used to probe the system where the photons are inelastically scattered by the magnons and are analyzed using a tandem Fabry-Peròt interferometer (TFPI) \cite{Sebastian2015}. Both the pump and probe lasers are focused close to their respective diffraction limits onto the sample using a high numerical aperture (N.A. = 0.75) objective to allow excitation and detection of magnons with high wavevectors up to $k\simeq 10 ~\text{rad}/\mu\text{m}$. The setup is also equipped with a pair of galvo-mirrors and lenses providing the means to measure the SW amplitude at a relative distance \textit{l} between the pump and the probe beam spots. 
	
	The sample under study was a 20~nm thick poly-crystalline NiFe (permalloy) thin film deposited on a sapphire substrate. The permalloy film  was coated with 2~nm of dielectric SiOx to prevent surface oxidation. An oblique (82$^\circ$ out-of-plane) magnetic field ($H_{app}$) is applied as shown in Fig.\ref{fig:1}. 
	
	\begin{figure}
		\centering
		\includegraphics[width=\linewidth]{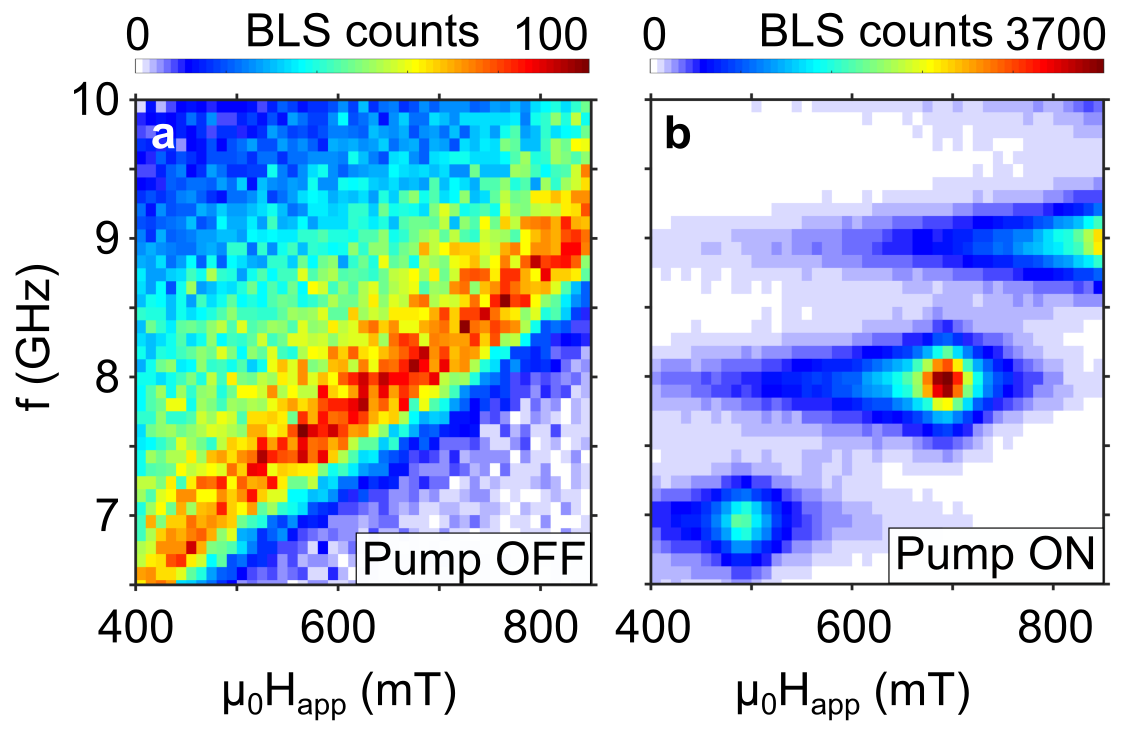}
		\caption{Field vs Frequency plot of the spin wave amplitude for NiFe 20 nm thin film (a) showing thermal magnons and (b) stimulated magnons by a pump laser with a fluence of 3.6~mJ/cm$^2$. }
		\label{fig:2}
	\end{figure}
	
	\emph{Results: }
	Fig.~\ref{fig:2} shows the  BLS spectra obtained for a small range of field from 400 to 850~mT with the pump (a) off and (b) on.  When the pump is off, the spectrum corresponds to the thermal magnons, which show a sharp field-dependent cut-off towards lower frequencies, corresponding to the SW gap, and a more gradual decrease towards higher fields due to the drop-off in wave vector resolution of the BLS. When we turn on the pump laser, a series of peaks, corresponding to the harmonics of the 1 GHz repetition rate, appear, showing how additional magnons are strongly excited at frequencies matching the repetition rate. The peaks are approximately located at the onset of the broadband ferromagnetic resonance signal. Each individual harmonic mode at 7, 8 and 9~GHz in Fig.~\ref{fig:2}(b) from the peak are extended towards lower fields and drastically drop in intensity at higher fields. At low fields, these modes correspond to propagating spin waves with high $k$-vector; at higher fields the modes fall into the SW band gap and are evanescent in time and space (non-resonant, forced oscillations). 
	
	\begin{figure*}
		\centering
		\includegraphics[width=0.9\textwidth]{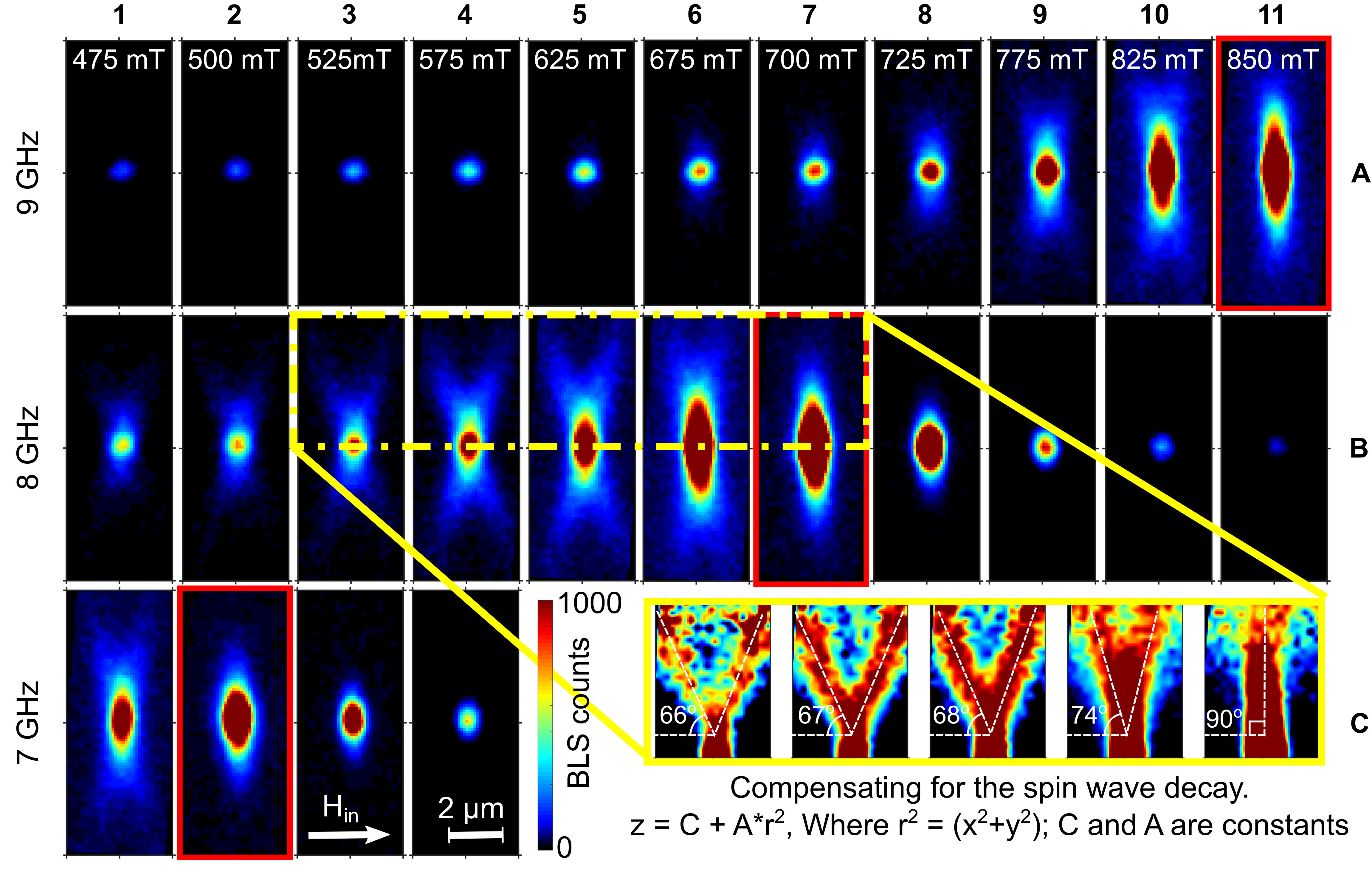}
		\caption{Spatial profiles of NiFe thin film measured in an area 4 x 10 $\mu m^2$ for 11 different field values. The area maps are plotted for 7, 8, and 9~GHz modes. The red outlined maps correspond to the maps plotted for the peak SW amplitude for individual modes in Fig.~\ref{fig:2}(b). The color scale is the same for all plots. The inset contains propagation characteristics of the 8 GHz mode for five  field values (B3-B7) giving an estimate of the caustic angle. The color maps in the inset are compensated for the spin wave decay.}
		\label{fig:3}
	\end{figure*}
	To determine the spatial extent of the SW modes at different applied field strength, we collected 11 SW intensity area maps for fields starting from 475~mT to 850~mT (Fig.~\ref{fig:3}). Since the BLS microscope separates counts directly in the frequency domain, the three modes (7, 8  and 9 GHz) can be separately plotted by selecting the specific channels corresponding to individual frequencies. All subplots use the same color scale, shown at the bottom of the figure. The spatial profiles highlighted by red outline (A-11, B-7 and C-2) correspond to the maximum SW intensity on the field sweep plot (Fig.~\ref{fig:2}(b)). They clearly show a strong straight extension along the direction perpendicular to the in-plane component of the applied field ($H_{in}$). At higher fields the spatial profile transforms into a localized spot which eventually fades away as the now evanescent SW mode moves deeper into the SW gap. 
	
	Reducing the field, on the other hand, dramatically changes the spatial profile at all three frequencies. We here focus on the 8 GHz mode where fields lower than 700~mT show a prominent caustic behavior with a propagation angle that decreases with field (\emph{i.e.}~with increasing $k$-vector); this can be seen in the highlighted yellow section. Here the dashed white lines are a guide to the eye to follow how the SW beam pattern changes from  perpendicular propagation \emph{w.r.t.}~the in-plane component of the applied field to a propagation angle of $\sim66^{\circ}$ with decreasing field. The weaker SW intensity in subplots B1 and B2 of Fig.~\ref{fig:3} are due the decreasing efficiency of both the excitation and the BLS detection at higher wavevector. The same overall behavior is found at 7 and 9 GHz.  
	
	\emph{Theoretical analysis}: It is convenient to start with considering the anisotropic properties in Fourier space of a continuous medium. A small source with the characteristic size $r$ can emit waves with all wavevectors $(\mathbf{k})$ up to $\sim 1/r$ and group velocities $(\mathbf{v}_{gr})$, defined by a dispersion relation of the media as $\mathbf{v}_{gr}= 2 \pi \nabla f (\mathbf{k})$. Thus, if the surface of the emitted frequency $f_0 (\mathbf{k})$ (isofrequency surface) contains ``flat" regions, all waves with wavevectors in this region have the same direction of $\mathbf{v}_{gr}$, which leads to the formation of high-intensity unidirectional beams \cite{davies2015generation, gieniusz2013single}. The analysis of caustic SW beams in a thin magnetic film is described in detail in \cite{Schneider2010}, which we follow here. First, we calculate the magnon dispersion following a simplified version of Eq. (45) in    \cite{kalinikos1986theory}. For in-plane SWs, this reads: 
	\begin{align*}
	f &= \frac{\gamma \mu_0}{2\pi}\sqrt{(H_{int}+M_s l_{ex}^2 k^2)(H_{int}+M_s l_{ex}^2 k^2+M_s F_0)},\\[10pt]
	\vspace{25mm}
	F_0 &= P_0+\cos^2\theta_{int}\left[1-P_0(1+\cos^2\phi)\right.+ \numberthis \label{eq:disp}\\
	&+M_s\left.\left(P_0(1-P_0)\sin^2\phi\right)/\left(H_{int}+M_s l_{ex}^2k^2\right)\right],
	\end{align*}
	where $P_0 = 1-(1-\exp(-kL))$, $l_{ex}= \sqrt{2 A/(\mu_0 M_s^2)}$, and $H_{int}$ and $\theta_{int}$ define the value and OOP angle of the internal field; these can be found using the solution of the magnetostatic problem (e.g. Eq.(2.3) in   \cite{houshang2018spin}):
	\begin{eqnarray}
	H_{app}\cos \theta_{app} &=& H_{\mathrm{int}} \cos \theta_{\mathrm{int}} \\
	H_{app}\sin \theta_{app} &=& (H_{\mathrm{int}} + M_{\mathrm{s}}) \sin \theta_{\mathrm{int}},
	\label{eq:theta}
	\end{eqnarray} 
	where $H_{app}$ denotes the external magnetic field applied at an OOP angle $\theta_{app}$.
	
	The following parameters were used: saturation magnetization $M_s=0.98~\text{T}$, exchange stiffness $A=11.3~\text{pJ/m}$, film thickness $L=20$~nm, and $\theta_{app}=82^{\circ}$. The direction between the in-plane component of the field, $H_{in}$, and the wavevector, $\mathbf{k}$, is defined by $\phi$. The frequency of magnons vs.~wavevector components $k_x = k\sin\phi$ and $k_y= k\cos\phi$ is shown by contour plots in Fig.4. 
	
	In the next step we extract the isofrequency line $k_x=g(f, k_y)$ for the frequency of excitation $f=8~\text{GHz}$, which is shown by the red line in Fig.4. The condition for a caustic beam is a ``flat" region on this curve, namely:
	\begin{equation}
	\partial ^2 g(f, k_y)/\partial k_y^2=0
	\label{eq:condition}
	\end{equation}
	from which we can extract the wavevectors of caustic beams $\mathbf{k}_c$, which are shown on Fig.~4a and 4b by the green dots. The group velocities and, thus, the propagation directions of the caustic beams are perpendicular to the isofrequency line at these points, which can be written as $\pi/2 - \phi_c=\arctan(\partial g(f, k_y)/\partial k_y)$. The directions of the propagation are shown by green arrows.

	\begin{figure}
		\centering
		\includegraphics[width=0.48\textwidth]{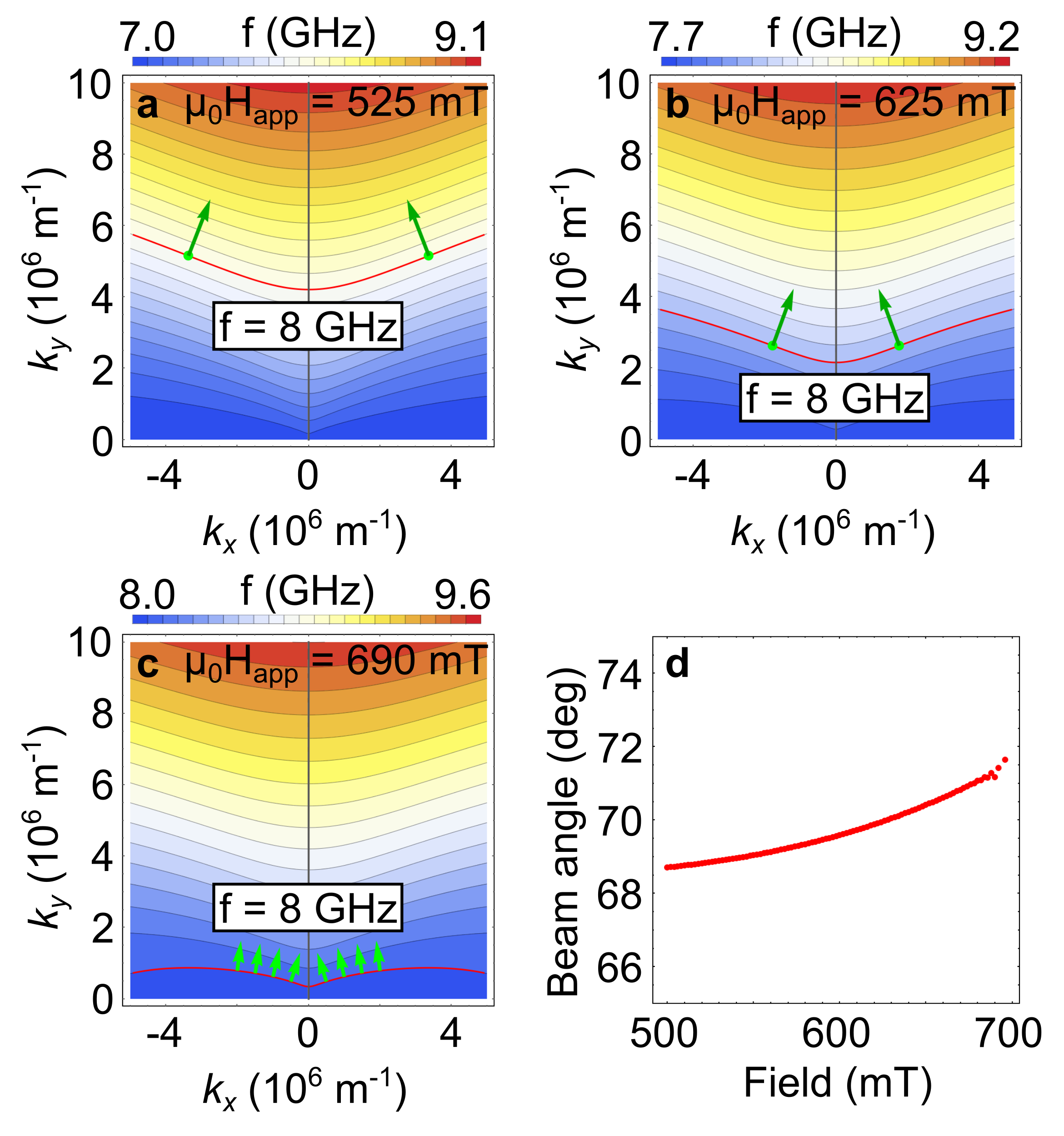}
		\caption{a) -- c) Isofrequency curves for 8 GHz for different applied fields. Caustic points defined by Eq. (4) are shown by green dots, while directions of the caustic beams by the darker arrows. c) shows the preffered direction of the emited SWs for the high field, where the caustic X-shape pattern doesn't exist; d) Caustic beam angle vs applied magnetic field.}
		\label{fig:4}
	\end{figure}
	
	With increasing field, the wavevector of the caustic beam decreases (see Fig.~4d), until at some point $k_x \to 0$. At this point, the condition in Eq.~(\ref{eq:condition}) can no longer be satisfied for the particular isofrequency curve, and, as a consequence, the caustic beam no longer exists. Instead the isofrequency shows a non-zero second derivative everywhere as illustrated by the green arrows in Fig.~4c. For our choice of parameters, this happens at $H_{app}\simeq 0.69~\text{T}$. 
	One can notice that the slope of isofrequency curve at high fields is quite small. Taking into account its symmetry for positive and negative $k_x$, this leads to the focusing of the emitted SWs primarily into the $y$ direction, as we observed above from the BLS spatial maps.
	
	In conclusion, we have demonstrated the generation of field-tunable caustic SW beams in 20 nm NiFe films using an all-optical excitation mechanism based on a femtosecond laser comb. The spatial profiles and frequency content of the generated SWs were mapped out using scanning Brillouin Light Scattering microscopy. With a high repetition rate laser, we excite only those SWs whose frequency matches the harmonics of the fs laser 1 GHz repetition rate. The implications of suppressing all other spin wave modes other than the one corresponding to the harmonic modes of the pump laser leads to the possibility of controlling the emission direction by tuning the magnetic field. Theoretically, we have calculated the angle of the caustic SW beams with reference to the in-plane component of the applied field by plotting the iso-frequency lines and applying the condition for caustic beams. Experimentally, we have plotted the area maps of the spin wave amplitude at various magnetic field strength values, and the estimated approximate values of the caustic beam angles agree well with theory. Additionally, by tuning the field, we are able to change the propagation characteristics from localized at high fields, to strongly propagating in a single direction when the ferromagnetic resonance frequency matches the harmonics of pump laser, and to X-shaped caustic patterns at lower fields for a particular SW mode. The highly tunable properties of the excited SWs can be utilized in the field of photomagnonics. 
	
	\emph{Acknowledgments} Early discussions with R.~E.~Camley are gratefully acknowledged. This work was partially supported by the Swedish Research Council (VR), the Knut and Alice Wallenberg foundation (KAW), and the Horizon 2020 research and innovation programme (ERC Advanced Grant No.~835068 "TOPSPIN").

\end{document}